\documentclass[12pt,a4paper]{article}
\usepackage{fullpage}
\usepackage{newcite}
\usepackage{amsbsy}
\usepackage{amsfonts}
\usepackage{amssymb}
\usepackage{newurl}
\usepackage[dvips]{graphicx}
\usepackage{subfigure}
\begin{document}

\begin{flushright}
\textsf{21 February 2003}
\\
\textsf{hep-ph/0302173}
\end{flushright}

\vspace{1cm}

\begin{center}
\large
\textbf{Last CPT-Invariant Hope for LSND Neutrino Oscillations}
\normalsize
\\[0.5cm]
\large
C. Giunti
\normalsize
\\[0.5cm]
INFN, Sezione di Torino, and Dipartimento di Fisica Teorica,
\\
Universit\`a di Torino,
Via P. Giuria 1, I--10125 Torino, Italy
\\[0.5cm]
\begin{minipage}[t]{0.8\textwidth}
\begin{center}
\textbf{Abstract}
\end{center}
It is shown that the 99\% confidence limits from the analyses of
the data of cosmological and neutrino experiments
imply a small marginally allowed region
in the space of the neutrino oscillation parameters
of $3+1$ four-neutrino mixing schemes.
This region
can be confirmed or falsified by experiments
in the near future.
\end{minipage}
\end{center}

The impressive results of the first year of observations
of the
Wilkinson Microwave Anisotropy Probe (WMAP)
\cite{Bennett:2003bz}
combined with other recent high precision cosmological data sets
(2dF Galaxy Redshift Survey,
Lyman $\alpha$ forest data,
CBI and ACBAR
microwave background data;
see references in Ref.~\cite{Spergel:2003cb})
allowed to derive tight constraints on the values of cosmological
parameters
\cite{Spergel:2003cb}.
In particular,
the combined fit of cosmological data
performed in Ref.~\cite{Spergel:2003cb}
imply the tight bound
\begin{equation}
\Omega_\nu h^2 < 7.6 \times 10^{-3}
\qquad
\mbox{(95\% confidence limit)}
\label{002}
\end{equation}
for the contribution of massive neutrinos to the energy density
of the universe
($\Omega_\nu$ is the neutrino density relative to the critical density,
$h$ is the Hubble constant in units of
100 km/s/Mpc),
whose dependence on the values of the neutrino masses
$m_i$
is given by
\cite{Gershtein:1966gg,Cowsik:1972gh}
\begin{equation}
\Omega_\nu h^2
=
\frac{ \sum_i m_i }{ 93.5 \, \mathrm{eV} }
\,.
\label{001}
\end{equation}

Strong evidences in favor of neutrino oscillations 
have been obtained in
atmospheric
\cite{Fukuda:1998mi,%
Allison:1999ms,%
Ambrosio:2000qy},
solar
\cite{Cleveland:1998nv,%
Hampel:1998xg,%
Altmann:2000ft,%
Abdurashitov:2002nt,%
Fukuda:2002pe,%
Ahmad:2002jz},
long-baseline reactor
\cite{hep-ex-0212021}
and accelerator
\cite{Ahn:2002up}
neutrino experiments.
The results of all these neutrino experiments
can be explained in the framework of standard
three-neutrino mixing
(see Refs.~\cite{BGG-review-98,Gonzalez-Garcia:2002dz})
with the two squared-mass differences
\begin{equation}
\Delta{m}^2_{\mathrm{sol}}
\simeq
7 \times 10^{-5} \, \mathrm{eV}^2
\,,
\qquad
\Delta{m}^2_{\mathrm{atm}}
\simeq
2.5 \times 10^{-3} \, \mathrm{eV}^2
\,,
\label{003}
\end{equation}
which generate, respectively,
solar and atmospheric neutrino oscillations.
In this case,
the bound (\ref{002}) can be saturated only if
the three neutrino masses are almost degenerate
(see Ref.~\cite{Bilenky:2002aw}),
and
each neutrino mass is bounded by \cite{Spergel:2003cb}
\begin{equation}
m_i < 0.23 \, \mathrm{eV}
\qquad
\mbox{(95\% confidence limit)}
\,.
\label{004}
\end{equation}

Neutrino oscillations
in the framework of three-neutrino mixing cannot explain simultaneously
the results of
solar, atmospheric, long-baseline neutrino experiments
and
the evidence in favor of short-baseline
$\bar\nu_\mu \to \bar\nu_e$
transitions found in the LSND experiment
\cite{Aguilar:2001ty},
because
an additional large squared-mass difference,
\begin{equation}
\Delta{m}^2_{\mathrm{LSND}}
\sim
1 \, \mathrm{eV}^2
\,,
\label{005}
\end{equation}
is needed.
Taking into account only the need of
three independent squared-mass differences,
the minimal framework that may be able to
explain all data with neutrino oscillations
is four-neutrino mixing
(see Refs.~\cite{BGG-review-98,Gonzalez-Garcia:2002dz}).
There are two types of schemes
that provide the needed hierarchy of squared-mass differences in Eqs.~(\ref{003}) and (\ref{005}):
$2+2$ schemes with two pairs of almost degenerate
massive neutrinos separated by the so-called ``LSND gap'' of the order of 1 eV,
and
$3+1$ schemes with a triplet of almost degenerate
massive neutrinos and an isolated massive neutrino separated by the LSND gap.
In spite of the apparent large freedom in four-neutrino mixing schemes,
which have
three independent squared-mass differences and six mixing angles,
the present data of neutrino experiments are so rich that
four-neutrino schemes are tightly constrained.
In Refs.~\cite{Maltoni:2002xd,hep-ph-0209368}
it has been shown that $2+2$ schemes are strongly disfavored by data
and $3+1$ schemes fail to provide an acceptable fit of the data,
except for a few marginally acceptable
regions in parameter space.

\begin{figure}[t]
\begin{center}
\null
\hfill
\subfigure[
``Normal''
]{
\includegraphics*[bb=182 460 356 773, width=0.2\textwidth]{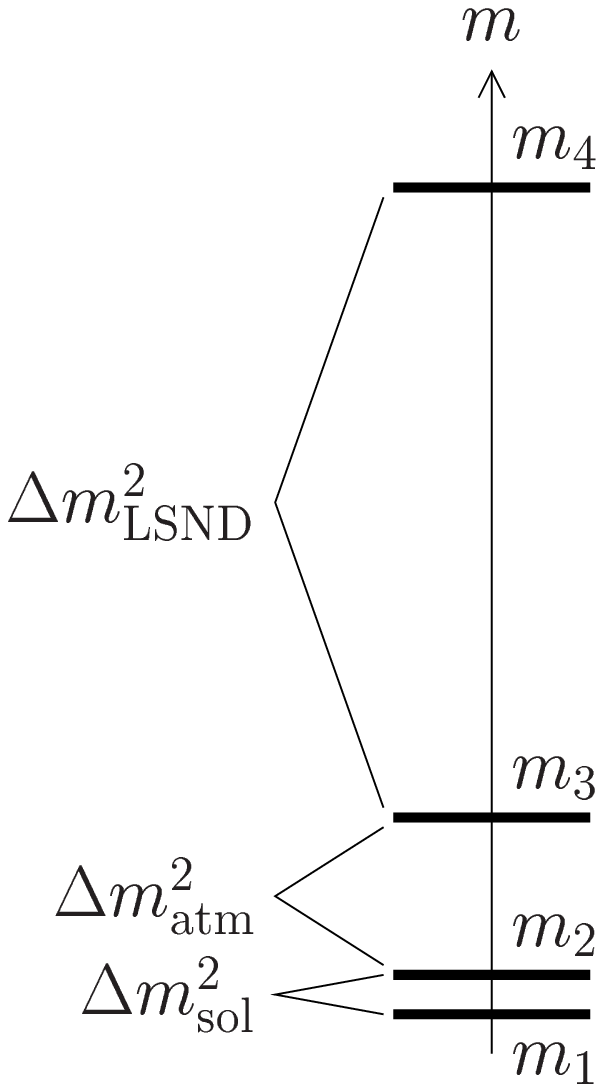}
\label{31n}
}
\hfill
\subfigure[
``Inverted''
]{
\includegraphics*[bb=182 460 356 773, width=0.2\textwidth]{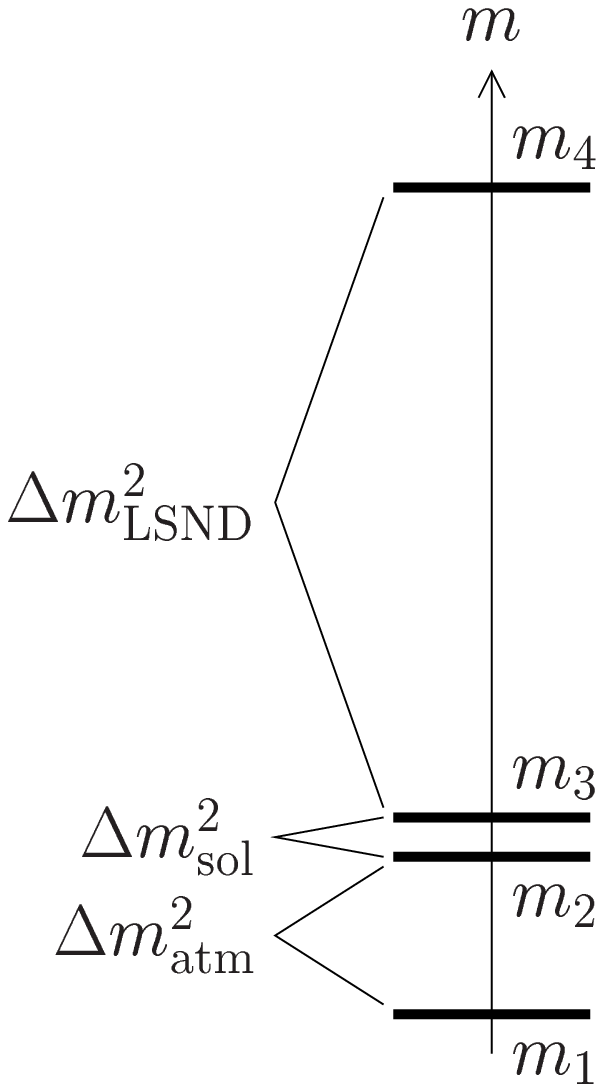}
\label{31i}
}
\hfill
\null
\end{center}
\caption{ \label{31}
Schematic view of the masses in the two $3+1$ four-neutrino schemes
in which the isolated neutrino is the heavier one.
}
\end{figure}

\begin{figure}[t]
\begin{center}
\includegraphics*[width=0.7\textwidth]{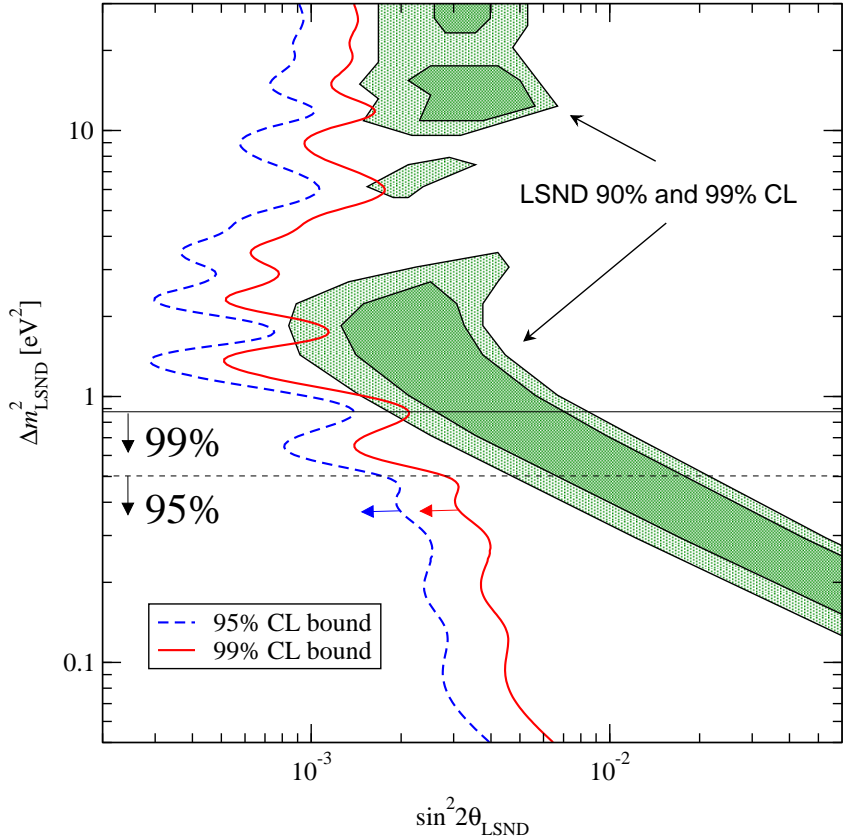}
\end{center}
\caption{ \label{bounds}
Figure taken from Ref.~\cite{hep-ph-0209368},
on which we have superimposed the
95\% confidence limit in Eq.~(\ref{95-2}) (dashed horizontal line)
and the
99\% confidence limit in Eq.~(\ref{99-2}) (solid horizontal line).
}
\end{figure}

In a very interesting paper
\cite{hep-ph-0302131}
Pierce and Murayama
considered the marginally allowed $3+1$ schemes
shown schematically in Fig.~\ref{31},
in which the isolated neutrino is the heavier one,
$\nu_4$ with mass $m_4$.
In this case,
only the heaviest massive neutrino
saturates the cosmological bound in Eq.~(\ref{002}),
leading to the limit
\begin{equation}
m_4 < 0.71 \, \mathrm{eV}
\qquad
\mbox{(95\% confidence limit)}
\,,
\label{95-1}
\end{equation}
which straightforwardly implies
\begin{equation}
\Delta{m}^2_{\mathrm{LSND}} < 0.50 \, \mathrm{eV}^2
\qquad
\mbox{(95\% confidence limit)}
\,.
\label{95-2}
\end{equation}
In Fig.~\ref{bounds}
we show this limit (dashed horizontal line)
superimposed to Fig.~4 of Ref.~\cite{hep-ph-0209368},
which shows the LSND-allowed regions at 90\% and 99\% C.L.
in the plane of the mixing parameters
$\sin^22\theta_{\mathrm{LSND}}$, $\Delta{m}^2_{\mathrm{LSND}}$,
compared with the upper bound for $\sin^22\theta_{\mathrm{LSND}}$
from short-baseline, atmospheric and solar neutrino data.
These bounds are due to the fact that
in $3+1$ four-neutrino mixing schemes
the oscillation amplitude
$\sin^22\theta_{\mathrm{LSND}}$
in the LSND experiment
(as well as in other short-baseline
$\bar\nu_\mu\to\bar\nu_e$
and
$\nu_\mu\to\nu_e$
experiments)
is given by
\begin{equation}
\sin^22\theta_{\mathrm{LSND}}
=
4 |U_{e4}|^2 |U_{\mu4}|^2
\,,
\label{006}
\end{equation}
where $U$ is the unitary mixing matrix
(see Ref.~\cite{BGG-review-98}).
The values of $|U_{e4}|^2$ and $|U_{\mu4}|^2$
are constrained by the negative results
of short-baseline reactor
$\bar\nu_e\to\bar\nu_e$
and accelerator
$\nu_\mu\to\nu_\mu$
disappearance experiments,
whose oscillation amplitudes are given,
respectively,
by
\begin{equation}
\sin^22\theta_{\bar\nu_e\to\bar\nu_e}
=
4 |U_{e4}|^2 \left( 1 - |U_{e4}|^2 \right)
\,,
\qquad
\sin^22\theta_{\nu_\mu\to\nu_\mu}
=
4 |U_{\mu4}|^2 \left( 1 - |U_{\mu4}|^2 \right)
\,.
\label{007}
\end{equation}
Large values of $|U_{\mu4}|^2$
are forbidden by the results of atmospheric neutrino experiments,
because in order to have $\nu_\mu$ oscillations
due to
$\Delta{m}^2_{\mathrm{atm}}$
the muon neutrino must have a large mixing with the three light massive neutrinos,
which by unitarity implies a small mixing with the heavy isolated massive neutrino.
Similarly,
large values of $|U_{e4}|^2$ are forbidden by the results of
solar neutrino experiments.
The resulting upper bounds on $\sin^22\theta_{\mathrm{LSND}}$
calculated in Ref.~\cite{hep-ph-0209368}
at 95\% and 99\% C.L.
are represented by the wiggling lines in Fig.~\ref{bounds}.

One can see from Fig.~\ref{bounds}
that the cosmological upper bound
(\ref{95-2})
excludes all the regions which are marginally allowed
by solar, atmospheric and terrestrial neutrino experiments.
From this comparison 
Pierce and Murayama
\cite{hep-ph-0302131}
concluded that a four-neutrino mixing explanation of the LSND results
is strongly disfavored
and ``the only way to reconcile LSND with the cosmological data
is to have CPT violation''
\cite{Murayama:2000hm,hep-ph-0212116}.

Here we would like to point out that,
although there is a clear tension between different data
interpreted in the framework of four-neutrino mixing,
before considering this possibility ruled out
we should apply more caution.
A way to avoid premature conclusions
is to consider experimental limits with high confidence level
or high probability,
which is especially recommended for the comparison of results of different
experiments.
In this spirit,
the 95\% confidence of the limit in Eq.~(\ref{95-2})
does not appear to be sufficient to rule out four-neutrino mixing.

Unfortunately,
the authors Ref.~\cite{Spergel:2003cb}
did not give the values of more robust confidence limits.
However,
it is possible to extract additional information from Fig.~14 of their paper,
which shows the
cumulative probability of $\Omega_\nu h^2$.
One can see that the cumulative probability of $\Omega_\nu h^2$
is rather flat above the limit in Eq.~(\ref{002})
and requiring 99\% confidence may raise significantly the bound.

The precise values of the
cumulative probability of $\Omega_\nu h^2$
extracted from the Postscript file of Fig.~14 of Ref.~\cite{Spergel:2003cb}
are available at
\url{ftp://wftp.to.infn.it/pub/giunti/wmap-0302209/wmap-0302209-f14.txt}.
The plot in Fig.~14 of Ref.~\cite{Spergel:2003cb}
ends at
$\Omega_\nu h^2 = 0.010$,
where we find a cumulative probability of $1.4 \times 10^{-2}$
which correspond to a 98.6\% confidence limit.
For simplicity, let us approximate it with a 99\% confidence limit.
Hence,
in the $3+1$ schemes in Fig.~\ref{31} we have the bound
\begin{equation}
m_4 < 0.94 \, \mathrm{eV}
\qquad
\mbox{(99\% confidence limit)}
\,,
\label{99-1}
\end{equation}
and
\begin{equation}
\Delta{m}^2_{\mathrm{LSND}} < 0.87 \, \mathrm{eV}^2
\qquad
\mbox{(99\% confidence limit)}
\,.
\label{99-2}
\end{equation}
This bound is represented by the solid horizontal line in Fig.~\ref{bounds}.

From Fig.~\ref{bounds} one can see that the three 99\% confidence regions
obtained from the LSND results,
from the results of solar, atmospheric and terrestrial neutrino experiments,
and from cosmological data
overlap on a tiny region in the
$\sin^22\theta_{\mathrm{LSND}}$--$\Delta{m}^2_{\mathrm{LSND}}$
plane at\footnote{
See also the interesting Addendum 4 in Version 4 of the \texttt{hep-ph} version
of Ref.~\cite{Strumia:2002fw},
which appeared after completion of this paper. 
}
\begin{equation}
\sin^22\theta_{\mathrm{LSND}} \simeq 2.0 \times 10^{-3}
\,,
\qquad
\Delta{m}^2_{\mathrm{LSND}} \simeq 0.86 \, \mathrm{eV}^2
\,.
\label{allowed}
\end{equation}
This region corresponds to the region R2 found in Ref.~\cite{Giunti:2000ur}.

Although we have to admit that this region is only marginally allowed
and its probability is rather low,
we think that it is still not excluded and may constitute the
``last CPT-invariant hope for LSND neutrino oscillations''\footnote{
We do not consider here the constraints imposed by
Big-Bang Nucleosynthesis,
whose status is still controversial
(see Refs.~\cite{Dolgov:2002wy,Kainulainen:2002pu,astro-ph-0302431,astro-ph-0302433}).
}.

The predictions of the values of neutrino oscillation parameters in Eq.~(\ref{allowed})
for future experiments are rather strong.
Of course, such region should be measured
in the MiniBooNE experiment \cite{MiniBooNE-Nu2002},
which is aimed to check the LSND result.
Since this region is just below the upper bound
for $\sin^22\theta_{\mathrm{LSND}}$
from short-baseline, atmospheric and solar neutrino data,
for a value of
$\Delta{m}^2_{\mathrm{LSND}}$
where such bound is due to
the combined results of the
Bugey reactor $\bar\nu_e$ disappearance experiment \cite{Declais:1995su}
and the
CDHS accelerator $\nu_\mu$ disappearance experiment \cite{CDHS},
future experiments should find
$\bar\nu_e$ and $\nu_\mu$ disappearance at the borders of the
Bugey and CDHS bounds at
$\Delta{m}^2 \simeq 0.86 \, \mathrm{eV}^2$:
\begin{equation}
\sin^22\theta_{\bar\nu_e\to\bar\nu_e}
\simeq
7 \times 10^{-2}
\,,
\qquad
\sin^22\theta_{\nu_\mu\to\nu_\mu}
\simeq
0.1
\,,
\label{009}
\end{equation}
which correspond to
\begin{equation}
|U_{e4}|^2 \simeq 2 \times 10^{-2}
\,,
\qquad
|U_{\mu4}|^2 \simeq 3 \times 10^{-2}
\label{008}
\end{equation}
(see also Figs.~3--6 of Ref.~\cite{Giunti:2000ur}).
Furthermore,
if there are no fine-tuned cancellations
among the contributions to the effective Majorana mass
in neutrinoless double-$\beta$ decay
\cite{Giunti:1999jw},
this effective mass could be as high as
\begin{equation}
|U_{e4}|^2 m_4 \simeq 2 \times 10^{-2} \, \mathrm{eV}
\,,
\label{010}
\end{equation}
and could be observed in future experiments
\cite{Zdesenko:2001ee,Aalseth:2002sy,hep-ph/0206249,hep-ex-0302021}.
The contribution of the heavy neutrino $\nu_4$
to the effective neutrino mass in tritium beta decay
(see Ref.~\cite{Bilenky:2002aw})
is
\begin{equation}
|U_{e4}| m_4 \simeq 0.1 \, \mathrm{eV}
\,,
\label{011}
\end{equation}
which is not much below the 0.35 eV expected sensitivity of
the KATRIN experiment \cite{Osipowicz:2001sq}
and may be observed in the future.

In conclusion, we would like to remark that,
although the alternative possibility of large CPT violations
in the neutrino sector is undoubtedly very interesting,
great caution should be exercised before ruling out
four-neutrino mixing schemes that may allow an appealing
explanation of all neutrino experiments in the framework of
local quantum field theory with the standard three flavor neutrinos
and an additional
very intriguing sterile neutrino
(see Ref.~\cite{Volkas:2001zb}).
Considering 99\% confidence limits from the analyses of
the data of cosmological and neutrino experiments
we found a small marginally allowed region for the oscillation parameters
in $3+1$ four-neutrino mixing schemes,
which can be confirmed or falsified by experiments
in the near future.

\end{document}